# Street-based Topological Representations and Analyses for Predicting Traffic Flow in GIS

Bin Jiang and Chengke Liu

Department of Land Surveying and Geo-informatics
The Hong Kong Polytechnic University, Hung Hom, Kowloon, Hong Kong
Email: bin.jiang@polyu.edu.hk

**Abstract** It is well received in the space syntax community that traffic flow is significantly correlated to a morphological property of streets, which are represented by axial lines, forming a so called axial map. The correlation co-efficient (R square value) approaches 0.8 and even a higher value according to the space syntax literature. In this paper, we study the same issue using the Hong Kong street network and the Hong Kong Annual Average Daily Traffic (AADT) datasets, and find surprisingly that street-based topological representations (or street-street topologies) tend to be better representations than the axial map. In other words, vehicle flow is correlated to a morphological property of streets better than that of axial lines. Based on the finding, we suggest the street-based topological representations as an alternative GIS representation, and the topological analyses as a new analytical means for geographic knowledge discovery.

## 1. Introduction

Space syntax is developed as a tool for understanding spatial structure, and consequently human life (e.g., human movement) in space from a topological point of view (Hillier and Hanson 1984, Hillier 1996). Supported mainly by two empirical studies (Hillier et al. 1993, Penn et al. 1998), extended recently by Jiang (2007a), it is well received in the space syntax community that traffic flow (referring to either pedestrian or vehicle flow) is significantly correlated to a morphological property of streets (a strict definition of street will be given). The modeling process starts with representing streets as axial lines that are intersected forming an axial map, and then ranking the individual lines using graph-theoretic measures for prediction purposes. The axial line is the longest visibility line, representing a street or part of a street.

The axial representation has been questioned by researchers for its validity, as it is neither cognitively sound nor computationally operable (Jiang and Claramunt 2002, Ratti 2004). For example, a ring road often acts as a hub for traffic, and is represented by many axial lines, forming a close chain of lines. It is not cognitively sound, as the ring road is a well received cognitive entity in our minds. More critically, the axial map, consisting of the least number of the longest axial lines, can not be automatically generated, and must be drawn manually. Although the drawing process can be guided by a restricted rule (Hillier and Hanson 1984, Turner 2005), it is remarkably difficult in practice to guarantee that two axial maps by two different people (or one person at different times) are the same. This imposes a significant doubt on its validity for various applications. In this paper, we provide further evidence that an axial map is indeed not a good representation in predicting traffic flow. Instead an alternative street-based representation is suggested.

A street is defined as a linear geographic entity that stretches in two dimensional space, and is often given a unique name. It should not be confused to a street segment between two junctions. We will illustrate that traffic flow can be better predicted using street-based topological representations. The street-based topological representations are developed from the previous studies (Thomson 2003, Jiang and Claramunt 2004), where streets are used to replace axial lines for a kind of topological representation. In the context of his paper, the term topology, in contrast to geometry, refers to a description of the relationship between geographic objects or locations. At a GIS database level, streets are merged street segments which belong to the same name – named streets (Jiang and Claramunt 2004), or alternatively, streets are *naturally* merged street segments that form good continuity – natural streets (or strokes in terms of Thomson (2003)). Therefore we have two types of streets that constitute the street-based topological representations. The two representations are examined against observed traffic flow. It is found to our surprise that the two are better than the axial map for traffic forecast. This finding verifies that the street-based topological representations are a better alternative GIS representation, which has far reaching implications to applications of Geographic Information Systems (GIS) for traffic management and transport planning, and to the understanding of self-organized cities.

This paper can be put under another context of topological analysis, which forms the fundamental for the new science of networks (Newman, Barabási and Watts 2006). The new science of networks supports the study of real world networks of often large size, takes the view that the networks are not static, but evolving in a self-organized manner, and aims to understand complex behaviors of real world systems and their constituents from a topological perspective. The new science has received a disproportionate amount of attention in a variety of disciplines including for instance biology, physics, sociology and computer sciences. The new wave of research interest in real world networks is mainly triggered by two seminal papers (Watts and Strogatz 1998, Barabási and Albert 1999) published respectively in Nature and Science. It is found that many real world networks are far from random graphs (Erdös 1960), but demonstrate small world and scale free properties. The topological analysis has been applied to urban street networks for studying emerging properties (e.g. Jiang and Claramunt 2004, Rosvall et al. 2005, Buhl et al. 2006, Porta et al. 2006a, 2006b, Jiang 2007b). The topological analysis represents a new paradigm for geospatial analysis and modeling, with a focus on spatial interaction and relationships.

The contribution of this paper is three-fold: (1) it introduces topological representation in general and street-based topological representations in particular for structural analysis of urban systems; (2) it proves that street-based topological representations are superior to conventional axial maps via traffic prediction; (3) it provides a research prototype and related algorithms for further research along the line of topological analysis.

The remainder of this paper is structured as follows. In section 2, we introduce the street-based topological representations and topological measures for structural analysis, and speculate on the importance of the analysis. Section 3 reports in detail the experiments and results using the Hong Kong street network and Annual Average Daily Traffic (AADT) datasets, illustrating the fact that the street-based topological representations are superior to an axial map in traffic forecast. Finally section 4 concludes the paper.

## 2. Topological representations and analyses
In this section, we will introduce basic concepts of space syntax and some key measures for topological analysis in the context of GIS. The brief introduction is mainly for understanding this paper, and the reader is encouraged to consult relevant literature for more details (e.g. Hillier and Hanson 1984, Jiang et al. 2000, Jiang and Claramunt 2004).

### 2.1 Geometric versus topological representations of urban street networks
Space syntax consists of a set of spatial representations for analyzing urban environments at both the city and built levels (Jiang et al. 2000). We mainly in this paper concentrate on the methods for a city. To illustrate, let us use a fictional city as an example (Figure 1). This is a typical GIS representation with two layers of information: streets as a network layer and buildings as a polygon layer. The city consists of 7 streets: a ring road A, and another 6 streets (B-H), but they are represented as a network with 11 nodes and the corresponding links between them (Figure 2d). The street network (or more precisely, the space separated by the building blocks) can be represented by an axial map, in which 13 axial lines (A-M) are intersected at 18 nodes (1-18) (Figure 2a). The beauty of space syntax lies in the fact that it collapses an entire axial line as a node and line-line intersection as a link (Figure 2b). In general, a space syntax graph representing line-line relationship is a non-planar graph, while an axial map is a planar graph.

What we adopted for predicting traffic flow is the kind of graph that encodes a street-street relationship (Figure 2e). This figure and figure 2b both represent primal graphs for a line-line or street-street relationship. In contrast, dual graphs for a point-point relationship can be established. This was first suggested by Jiang and Claramunt (2002) in what is called a point-based space syntax. Both figures 2c and 2f represent dual relationships between points, with respect to their primal graphs shown in figure 2b and figure 2e. It has been concluded in the previous study (Jiang and Claramunt 2002) that the primal and dual representations are identical from the point of view of morphological analysis. In other words, the morphological properties of points along a line are the same as those of the line. In this respect, Batty (2004) provides an elegant mathematical description for exploring primal and dual relationships.

It is important to point out a major difference between axial line-based and street-based topological representations. An axial line (or visibility line) is an approximation of a linear space while walking in it, so it is perception-based. A street (represented by a central line) is cognitive-based, as it is given a unique name. The difference can be clearly seen in figures 2a and 2d. The lines and streets, both related but different in essence, form respectively the line-based and street-based topologies in terms of intersection. In other words, the topologies are an interconnected whole of lines or streets.

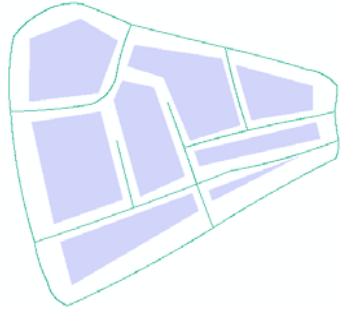

Figure 1: (color online) The map of a fictional city

| Geometric representations (planar graphs) | Topological representations (non-planar graphs) | |
|---|---|---|
| | Primal relation between lines | Dual relation between points |
| (a) Axial lines or axial map | (b) line-line relationship (13 axial lines) | (c) point-point relationship (18 points) |
| (d) Streets or street network | (e) street-street relationship (7 streets) | (f) point-point relationship (11 points) |

Figure 2: (color online) Geometric and topological representations

**2.2 Topological measures**

As shown previously, the line-line or street-street topologies are represented by a graph. In general, a graph ($G$) consists of a finite set of vertices (or nodes) $V = \{v_1, v_2, ... v_n\}$ (where the number of nodes is $n$) and a finite set of edges (or links) $E$, which is a subset of the Cartesian product $V \times V$. The graph can be represented as a matrix $R(G)$, whose element $r_{ij}$ is 1 if intersected, and 0 otherwise. Formally it is represented as follows:

$$R_{ij} = \begin{cases} 1 & \text{if i and j are intersected} \\ 0 & \text{otherwise} \end{cases} \quad (1)$$

It should be noted that this matrix *R(G)* is symmetric, i.e. $\forall r_{ij} \Rightarrow r_{ij} = r_{ji}$, and that all diagonal elements of *R(G)* are equal to zero. From a computational point of view, we compare each street to only those streets within the same envelop, a rectangular area that covers a street.

A range of measures for individual nodes (or vertices) are defined for topological analysis. Firstly the degree for a node is the number of other nodes directly connected to it. It is called connectivity in space syntax literature. Formally it is defined by:

$$M(v_i) = \sum_{j=1}^{n} R_{ij} \qquad (2)$$

Secondly, path length of a node is the measure of how far it is from all other nodes. It is defined by:

$$L(v_i) = \sum_{j=1}^{n} d(i,j), \qquad (3)$$

where $d(i,j)$ denotes the distance between two vertices *i* and *j*, which is the minimum length of the paths that connect the two vertices, i.e., the length of a *graph geodesic*.

Thirdly, clustering coefficient is the measure of the clustering degree of a node. It is defined as the probability that two neighbors of a given node are linked together, and measured by a ratio of the number of actual edges to that of possible edges.

$$C(v_i) = \frac{\# \text{ of actual edges}}{\# \text{ of possible edges}}, \qquad (4)$$

The measure path length can be used to derive integration, the key measure in space syntax. For the measure integration, it can be measured at both the local and global levels. To illustrate the difference, let us use the space syntax graph in figure 2b, but re-arranged all the nodes around node A (a sort of Bacon zero[*]) according to how far or close other nodes are from the node (Figure 3). Node A has a path length of 4 × 1 + 5 × 2 + 3 × 3 (which reads as 4 nodes in 1 step, 5 nodes in 2 steps, and 3 nodes in 3 steps), indicating how far it is from all other nodes. This is the global integration. Instead of all other nodes, if we consider nodes within two steps in equation (3), then we will have the path length at a local level, i.e., 4 × 1 + 5 × 2. This is the local integration. It is the default measure of space syntax for predicting traffic flow. For node A, its four neighbors would form 3 × 4/2 = 6 friendships, while there is only one actual friendship (between D and F), so the clustering coefficient of node A is 1/6.

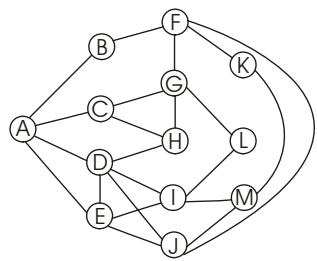

Figure 3: Adjusted graph showing a line-line relationship in Figure 2b

The above topological measures can be put into two categories according to how they are defined: local and global. For example, both connectivity and clustering coefficient are local measures, as only local nodes within a few steps (instead of all other nodes) are involved. In this connection, local integration can be considered to be a local measure. Global integration and path length are both global measures. Usually local measures are correlated to global measures, which gives a second order space syntax measure – intelligibility. Intelligibility is simply defined by the R square of the correlation. Connectivity, local and global integrations are key measures of space syntax. Degree (or connectivity), path length and clustering coefficient are three key measures for topological analysis. They constitute essential measures for exploring small world and scale free properties. In addition, Google's PageRank has emerged as an important measure for topological analysis (Jiang 2007a). For the sake of simplicity, we present a simplified version of local and global integrations. In fact, in the following experiments we adopted a normalized integration called Real Relative Asymmetry (refer to e.g. Jiang and Claramunt 2002 for more details) implemented in many space syntax software packages.

---

[*] For those who are unaware of Bacon number, actor or actress who has filmed with Kevin Bacon gets Bacon number one. Actor or actress, who has not filmed with Kevin Bacon, but has filmed with someone who has filmed with Kevin Bacon, gets Bacon number two. … This way, Kevin Bacon himself has the Bacon number zero. For more details, refer to the site http://en.wikipedia.org/wiki/Bacon_number.

### 2.3 Why the topological representations and analyses?
To illustrate why the topological representations and analyses matter, let us do a little experiment. Taking any street network shape file, merge all line segments between any pair of junctions as one single line, i.e., a street segment. This way we will get the network in which each line represents a segment between two adjacent junctions, which is truly a network of street segments. Based on the transformed network, we compute how many other segments intersected to any particular one. Figure 4a illustrates a histogram, which roughly follows a normal distribution with an exception at connectivity 3. However, a completely different pattern will emerge if we merge individual segments according to their unique street names. Figure 4b illustrates the histogram, where the connectivity ranges diversely from 1 to 60. As we can see, most streets have low degrees of connectivity, while a few have extremely high degrees. A more thorough investigation (Jiang 2007b) illustrates that 80% of streets have degrees less than the average, while 20% of streets have degrees higher than the average.

It is important to note that the links in the topological representations have no location sense, although it is derived from geographic objects or location. It sets a clear difference from the "topology" in the TIGER data model, developed by the U.S. Census Bureau in the 1960s. Coming back to the question raised, we can remark that topological representations and analyses help to uncover some hidden structure or patterns, which can not be illustrated by the geometric representation and analysis.

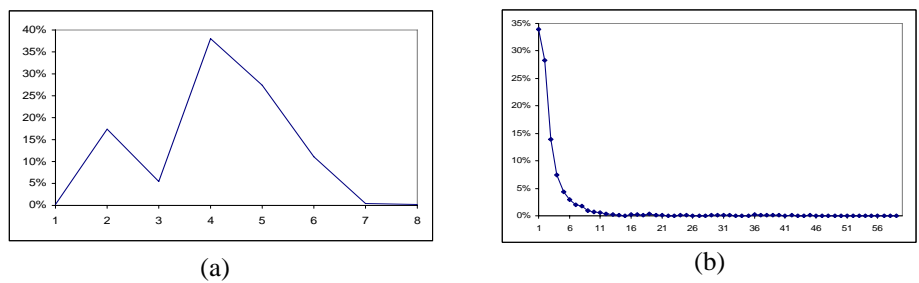

Figure 4: Degree distributions of street segments (a) and streets (b)

### 3. Experiments and results
In this section, we will use Hong Kong as a case study to illustrate that street-based topological representations are indeed superior to an axial map in predicting traffic flow. We also illustrate some topological properties of the Hong Kong street network. For the study, some essential data processing for forming topologies is carried out.

### 3.1 Data sources and algorithms for forming topologies
Two main data sources are used in the study. The first data source is the Hong Kong street network - central line street network. Using both the Hong Kong street network and building levels, we created an axial map consisting of 14378 axial lines. In order to generate street-based topological representations, we developed three algorithms for the processing. The first algorithm is based on the Gestalt principle of good continuity. The merging process can be described as follows (see also algorithm I in Appendix A): for every street segment, we trace its connected segments, and concatenate the segment and an adjacent one with the smallest deflection angle; this process will not be terminated until the smallest deflection angle reaches the threshold set. It should be noted that the threshold has a significant impact on the number of natural streets. We chose every 10 degrees between 20 and 70 as thresholds for determining continuity in the study. Thus for natural streets, we created 6 street-street topologies with different sizes (Table 1) for correlation comparison with traffic flow. The second algorithm is based on the name streets (see algorithm II in Appendix A). We merge the segments without names into neighboring segments using algorithm I, and then merge all segments according to unique names. It generates in total 7488 named streets. The reader may have noted a big difference in the number of natural streets and named streets. This is mainly because many Hong Kong streets have multiple lanes that are separate by barriers. Consequently, these streets are represented as many multiple central lines, although they are associated with the same name. However, they are treated as different streets in the process of generating natural streets. The third algorithm is to identify isolated streets or lines, and exclude them in forming topologies. It is pretty simple, and works as follows. Take one street or line as a root, and adopt the well-known Breadth-First Search algorithm to explore all those streets (or lines) that directly or indirectly connect to the root one. Those not connected either directly or indirectly to the root are isolated ones. The three algorithms have now integrated as the basic functionality of the newly released Axwoman 4.0 based on ArcGIS, and interested readers can refer to the site (www.hig.se/~bjg/Axowman) for more details.

The second data source is the Hong Kong Annual Traffic Census (Hong Kong Transport Department 2006), provided by the Traffic and Transport Survey Division of the Hong Kong Government Transport Department. The traffic census is conducted through a total of 3191 counting stations across the Hong Kong territory, but a significant portion of the stations function on a rotation basis. There were 829 counting stations functioning for the year 2005. We used the AADT generated from raw data collected from the counting stations, where inductive loops and pneumatic tubes are installed on a carriageway, and connected to the roadside automatic counters (Lee 1989). The counting stations were pinpointed onto a map in order to decide which axial lines or which streets (either named or natural) the stations belong to. We first roughly pinpoint, then adjust more precisely to locate them onto respective axial lines and streets. This pinpointing process is remarkably tedious, but we did it with help of a designed computer code. We also found that there are 4 counting stations that are impossible to pinpoint, thus are excluded from the study. Figure 5 illustrates the distribution of the 825 stations on the Hong Kong map.

Table 1: Size of axial line-based and street-based topologies

| Topology | Size |
| --- | --- |
| Axial lines | 14378 |
| Named streets | 7488 |
| Natural streets (20) | 21008 |
| Natural streets (30) | 17963 |
| Natural streets (40) | 15868 |
| Natural streets (50) | 14542 |
| Natural streets (60) | 13886 |
| Natural streets (70) | 13429 |

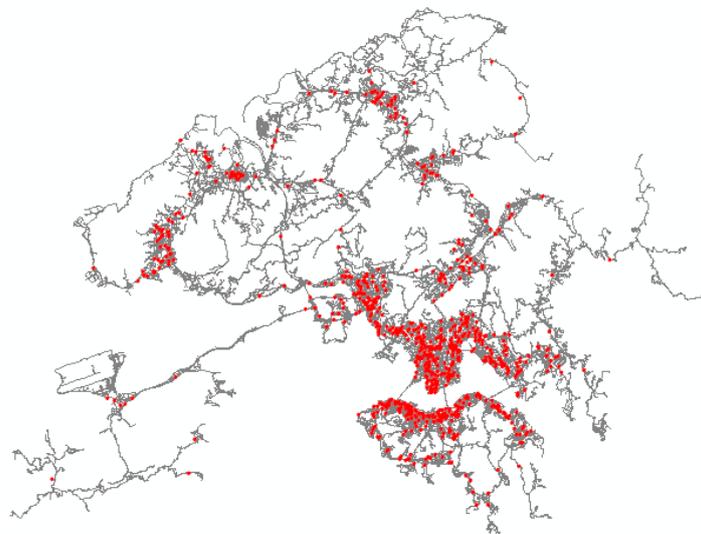

Figure 5: (color online) 825 counting stations across the Hong Kong territory

### 3.2 Computing topological properties

We applied equations (3) and (4) to get the measures for individual nodes of a topology. Then we take the average measures to assign to their respective topologies. The results are shown in table 2. We can remark that the average degree is around 3 or 4. However, for the axial line topology, its path length is bigger than its random counterpart. It appears that it may not be a small world, or have a rather weak small world property. On the other hand, both named street topology and natural street topology (created with a threshold of 60 degrees) are small worlds, since the path length L is pretty close to that of random counterparts $L_{rand}$. This finding of small worlds is not beyond our expectation, as it is a very common held property appearing in many real world network topologies. However, it reinforces our view speculated in section 2.3 as to how the topological representations help to uncover hidden structures and patterns.

Table 2: Small world parameters for three kinds of topologies
(NOTE: N = # of nodes, M = average degree of connectivity, L = path length, $L_{rand}$ = path length of random counterpart, C = clustering coefficient, $C_{rand}$ = clustering coefficient of random counterpart)

| Topology | N | M | L | $L_{rand}$ | C | $C_{rand}$ |
|---|---|---|---|---|---|---|
| Axial lines | 14378 | 2.76 | 33.88 | 9.43 | 0.08 | 0.0002 |
| Named streets | 7488 | 3.29 | 9.29 | 7.38 | 0.21 | 0.0005 |
| Natural streets (20) | 21008 | 3.79 | 23.42 | 7.47 | 0.42 | 0.0002 |
| Natural streets (30) | 17963 | 3.71 | 16.95 | 7.47 | 0.38 | 0.0002 |
| Natural streets (40) | 15868 | 3.65 | 12.91 | 7.47 | 0.35 | 0.0002 |
| Natural streets (50) | 14542 | 3.6 | 10.98 | 7.48 | 0.32 | 0.0002 |
| Natural streets (60) | 13886 | 3.57 | 10.12 | 7.5 | 0.31 | 0.0003 |
| Natural streets (70) | 13429 | 3.54 | 9.73 | 7.52 | 0.3 | 0.0003 |

Apart from small world property, the degree of axial lines and streets demonstrate a scale-free property, i.e., $p(x) \sim cx^{-\alpha}$. Figure 6 demonstrates log-log plots, whose x-axis and y-axis represent the logarithms of degree and cumulative probability. We can remark that the three log-log curves are pretty close to a straight line with an exponent around 2.0, thus a clear indication of scale free property. The scale-free property can be further described in detail as follows: about 80% of streets with a street network have length or degrees less than the average value of the network, while 20% of streets have length or degrees greater than the average. Out of the 20%, there are less than 1% of streets which can form a backbone of the street network (Jiang 2007b). Figure 7 highlights the top 1% of well connected streets, forming the backbone of the Hong Kong street network.

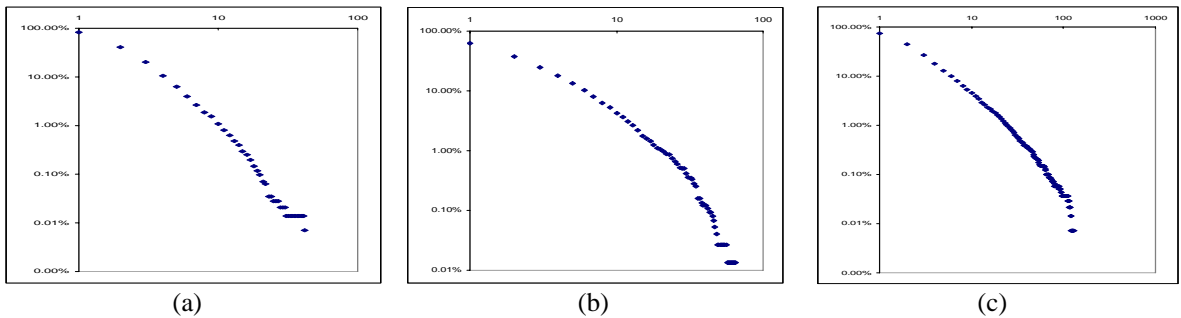

Figure 6: Power law distributions of axial lines (a), named streets (b) and natural streets (60) (c)

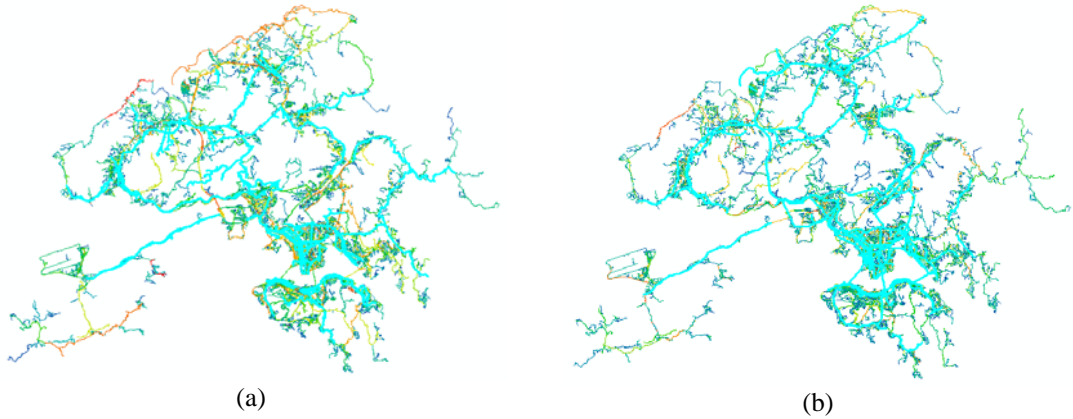

(a)                  (b)

Figure 7: (color online) 1% vital streets highlighted (turquoise) selected from (a) named streets, and (b) natural streets

The street-based topological representations play an important role in expanding our understanding of streets and a street network as a whole. Considering individual street segments as agents, the agents interact with their neighbors, forming what we call streets. The streets can be considered to be an emergence, whose size follows a power law distribution. In this respect, streets can be compared to avalanches of all sizes emerged from a sand pile, the model of self-organized criticality (Bak et al. 1987, Bak 1996). In other words, street segments are to streets what sand grains are to avalanches. The avalanches are emerged by gradually adding the sand grains in time dimension, whereas the streets are formed by gradually merging adjacent street segment in space dimension. The angle threshold we set for good continuity is sort of slope limit for forming avalanches in a sand pile. Therefore, streets or a street network in general can be understood as a self-organized phenomenon using the theory of self-organized criticality.

**3.2 Predictability of traffic flow**

We extracted those axial lines and streets that have counting stations for correlation comparison. Basically, flow observed in the counting stations along axial lines or streets is summed up to correlate to some topological measures (including degree, local integration and global integration) of the individual lines or streets (see Table 3). Unfortunately, the results are not particularly encouraging. As shown in table 3, R square value for the correlation between local integration and traffic flow is indeed the best for either axial lines or streets. We can observe that different thresholds for forming natural streets have some effect on the correlation between flow and topological measures. We note also from the experiments that natural streets using a threshold of 60 degrees seem to be the best option. In what follows, we will adopt this particular natural street-based topology for further analysis.

Table 3: Correlation coefficient (or R square value) between topological measures and traffic flow with different topologies
(NOTE: Connect = degree vs flow, LInteg = local integration vs flow, GInteg = global integration vs flow)

| Topology | Connect | LInteg | GInteg |
| --- | --- | --- | --- |
| Axial lines | 0.088 | 0.145 | 0.105 |
| Named streets | 0.164 | 0.277 | 0.275 |
| Natural streets (20) | 0.151 | 0.211 | 0.158 |
| Natural streets (30) | 0.187 | 0.258 | 0.206 |
| Natural streets (40) | 0.176 | 0.282 | 0.218 |
| Natural streets (50) | 0.166 | 0.314 | 0.242 |
| Natural streets (60) | 0.160 | 0.333 | 0.247 |
| Natural streets (70) | 0.142 | 0.308 | 0.218 |

The reason why the correlation is so poor can be attributed to the geographic nature of the Hong Kong territory. It consists of many islands, and much of the territory with mountains and hills remains undeveloped. The poor correlation could be possibly due to the uneven distribution of the counting stations in the sampled streets or axial lines. For this reason, we decided to take some small sample areas for a more in-depth study. We took two sampling methods. The first 10 samples are taken from the 10 most populated and urbanized areas (Figure 8a, refer to Appendix B for an enlarged view), while the second 9 areas are sampled according to different morphological patterns (Figure 8b, refer to Appendix C for an enlarged view). The morphological samples are put into three categories: grid-like, deformed-grid, and irregular (see Figure 9 for example). For all the samples, we chose local integration and correlate it with traffic flow using axial lines and streets. In what follows, we refer to the correlation coefficient between local integration and traffic flow as predictability.

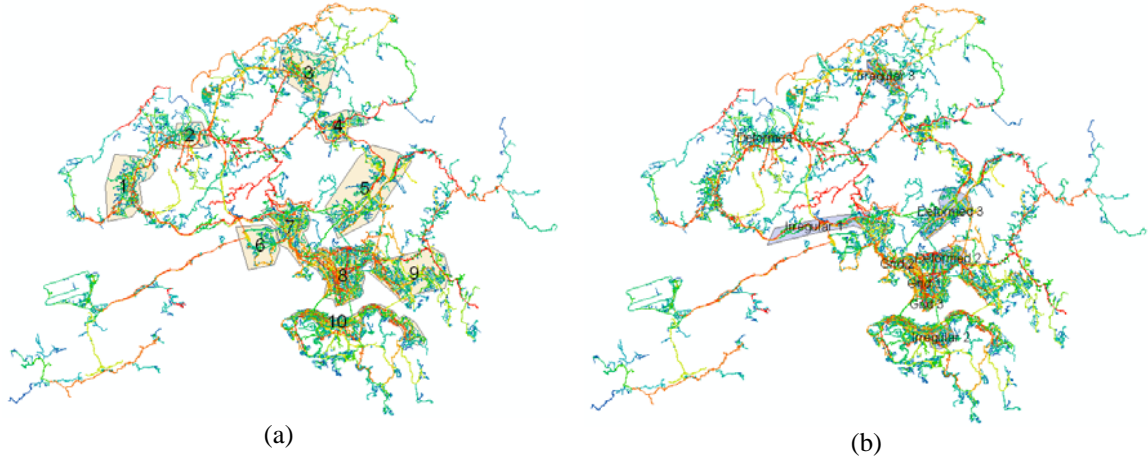

(a) (b)

Figure 8: (color online) Geographic distribution of sampled areas: (a) 10 most populated and intensively urbanized areas, and (b) 9 sampled areas according to morphology (Note: the network patterns of the sampled areas are shown at a detailed level in Appendix B)

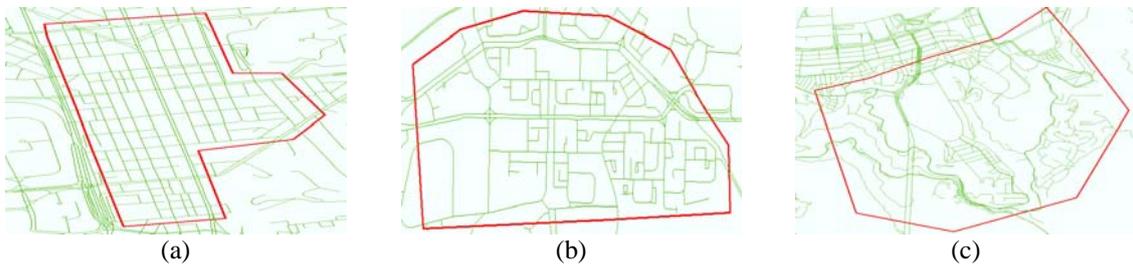

(a) (b) (c)

Figure 9: (color online) Grid-like (a), deformed-grid (b) and irregular (c) samples

The predictability for the first 10 samples is listed in table 4. We can note that the axial lines option is the poorest (second row with table 4), while both streets options show significant improvement (third and fourth rows with table 4) in predictability. We notice that the streets options demonstrate much better predictability than axial lines, either individually or overall by mean. The predictability for some areas goes up to 0.7.

Table 4: Predictability for traffic flow with the 10 sampled areas

| Areas | 1 | 2 | 3 | 4 | 5 | 6 | 7 | 8 | 9 | 10 | Mean |
|---|---|---|---|---|---|---|---|---|---|---|---|
| Axial lines | 0.17 | 0.00 | 0.04 | 0.06 | 0.11 | 0.00 | 0.12 | 0.13 | 0.18 | 0.06 | 0.09 |
| Named streets | 0.50 | 0.70 | 0.18 | 0.30 | 0.47 | 0.02 | 0.25 | 0.44 | 0.24 | 0.17 | 0.33 |
| Natural streets | 0.35 | 0.39 | 0.24 | 0.57 | 0.56 | 0.45 | 0.38 | 0.33 | 0.23 | 0.31 | 0.38 |

The predictability of axial lines is significantly influenced by the morphology or shape of areas (Table 5). It appears that the predictability of axial maps for grid-like areas is pretty good, which can be compared to that of streets. However the predictability of axial maps decreases dramatically for a deformed-grid and irregular areas. We can notice in the mean time that the predictability of both named and natural streets is rather stable, even with the morphological change. Overall, both named and natural streets have a better predictability than axial lines.

Table 5: Predictability for traffic flow with the 9 sampled areas according to morphology
(Note: G* = Grid-like samples, D* = Deformed-grid samples, I* = Irregular samples)

| Areas | G1 | G2 | G3 | D1 | D2 | D3 | I1 | I2 | I3 | Mean |
|---|---|---|---|---|---|---|---|---|---|---|
| Axial lines | 0.4 | 0.32 | 0.6 | 0.22 | 0.18 | 0.3 | 0.02 | 0.16 | 0.08 | 0.25 |
| Named streets | 0.43 | 0.59 | 0.57 | 0.58 | 0.62 | 0.57 | 0.39 | 0.44 | 0.43 | 0.49 |
| Natural streets | 0.35 | 0.69 | 0.61 | 0.56 | 0.6 | 0.61 | 0.55 | 0.56 | 0.41 | 0.53 |

We also computed intelligibility for the 9 sample areas, and grid-like areas, as it turns out, are more intelligible than irregular ones. This reconfirms the finding in the previous studies (e.g. Penn 2003). However, this finding has no significant similarity to streets either named or natural. We can easily notice that intelligibility does not change significantly when morphology changes from grid-like, through deformed grid to irregular. This may reconfirm that street-based topologies are indeed a better option for traffic forecast.

Table 6: Intelligibilities of the 9 sampled areas according to morphology
(Note: G* = Grid-like samples, D* = Deformed-grid samples, I* = Irregular samples)

| Areas | G1 | G2 | G3 | D1 | D2 | D3 | I1 | I2 | I3 | Mean |
|---|---|---|---|---|---|---|---|---|---|---|
| Axial lines | 0.59 | 0.56 | 0.50 | 0.02 | 0.23 | 0.13 | 0.12 | 0.17 | 0.01 | 0.26 |
| Named streets | 0.83 | 0.55 | 0.81 | 0.51 | 0.52 | 0.56 | 0.59 | 0.54 | 0.23 | 0.59 |
| Natural streets | 0.63 | 0.67 | 0.78 | 0.82 | 0.49 | 0.47 | 0.56 | 0.47 | 0.35 | 0.60 |

Throughout the above experiments, we have illustrated that (1) street-based topologies exhibit small-world and scale-free properties, while line-based topology shows a rather weak small world property; (2) the street-based representations are superior to conventional axial lines or the axial map in traffic forecast; (3) axial lines are significantly influenced by morphology in traffic forecast, while streets are much less so. The third point further reinforces the second point.

**4. Conclusion**

This paper introduced topological representations and analyses for predicting traffic flow in an urban environment. It is shown through experiments that the street-based topological representations and analyses are superior to the conventional axial map in predicting traffic flow. It is also shown that the topological representations, in comparison to conventional geometric oriented representation, can help to uncover some hidden structure or patterns as evident in our experiments. We believe that the street-based topological representations are better representations, not only for traffic prediction, but also for other well studied issues in space syntax such as pedestrian modeling, crime analysis, human wayfindings modeling etc. For instance, recently Tomko et al. (forthcoming) used named streets for exploring hierarchies of human spatial knowledge, proving that topologically prominent streets are indeed more likely to be known. As speculated briefly early in the text, axial lines tend to be perception-based, while street-based representations tend to be cognition-based. Based on this speculation, we can further conjecture that the axial line-based representation may be more suitable for pedestrian modeling, and street-based representation for vehicle prediction, because that the former is more visibility guided, thus local in nature, while the latter is more memory oriented, and global in nature.

From the point of view of geographic representation, street-based topological representations represent the highest abstraction of street networks, i.e., from the field-based representation, through object-based representation, to the street-based topological representations. The representations should become an alternative GIS representation, as they are illustrated to be both cognitively sound and computationally operable. Future work will concentrate on wider application studies based on the representations and analyses.

**Acknowledgements**
The work is financially supported by a Hong Kong Polytechnic University research grant. The traffic data are provided by the Traffic and Transport Survey Division of the Hong Kong Government Transport Department.

We would like further to thank the three referees and Itzhak Omer for providing some useful comments, and Wu Chen for his support in the course of the study.

**Appendix A: Algorithms for forming street-based topologies**

**Algorithm I (for extracting individual streets based on a good continuation)**

```
Input: Street segments-based shape file
Output: Street segments-based shape file with a new field namely StrokeID

While (not last segment)do
  Start with the first end point of the first segment in the attribute table of road
  if (the current segment is processed) then
     exit
  end
  Use a spatial filter to search for all segments intersected with that point
  Calculate the deflection angles between the found segments and the start one
  Get the minimums deflection angle and compare with the threshold
  if (the minimums deflection angle < threshold) then
     Concatenate that segment with the least angle with the start segment
     Change the status of that segment to processed
  end
  if (the route is traced to the end) then
      Start with the other end point of the current processing line to repeat the process
  end
  Pick out another unprocessed segment to continue the process
End while
```

**Algorithm II (for extracting individual streets based on a unique name)**
(Note: majority of the following lines are to assign unnamed segments into neighboring segments according to a good continuation)

```
Input: Street segments-based shape file with street names
Output: Street segments-based shape file with a new field namely NamedID

While (not last segment)do
 Start from the first unnamed segment (U)
 Search for connected streets to one side (e.g. left first) of segment U
 If (its neighbor streets have names) then
    Loop through all found segments to get the min. diversion angle (A1)
    Store the name of segment (N1) with angle A1
 Else
    Continue to search from the next connected unnamed segment
 End if
 Search for connected segment to the other side of segment U
 If (its neighbor streets have names) then
    Loop through all found segments to get the min. diversion angle (A2)
    Store the name of segment (N2) with angle A2
 Else
    Continue to search from the next connected unnamed segment
 End if
 If (A1<A2) AND (A1<60) then
    Assign the name of segment N1 to segment U
 Else If (A2<A1) AND (A2<60) then
    Assign the name of segment N2 to segment U
 Else
    Assign a random numeric number to segment U
 End if
 Pick out the next unprocessed unnamed segment
End while
```

Appendix B: (color online) 10 sampled areas according to districts

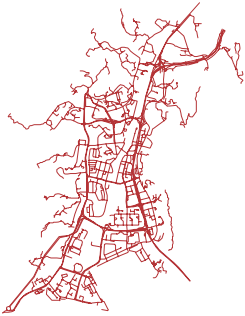
(a)

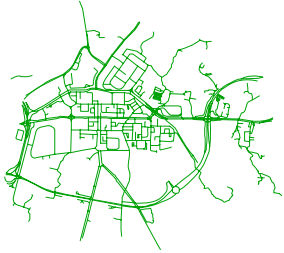
(b)

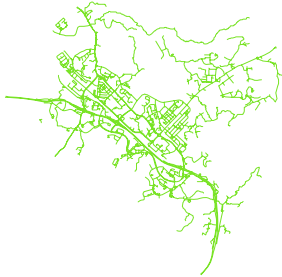
(c)

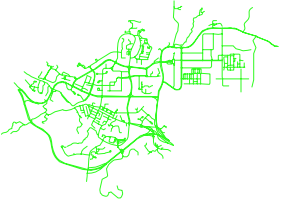
(d)

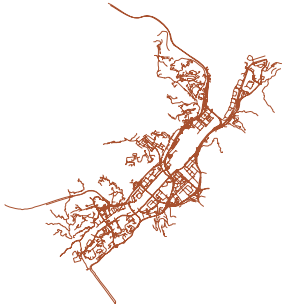
(e)

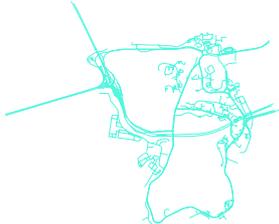
(f)

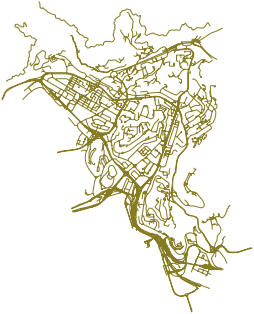
(g)

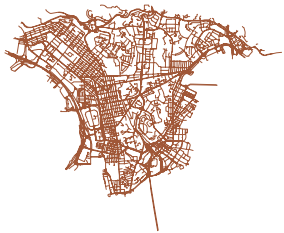
(h)

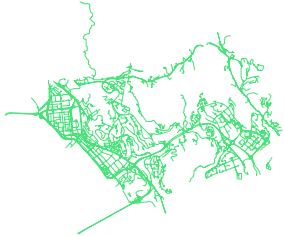
(i)

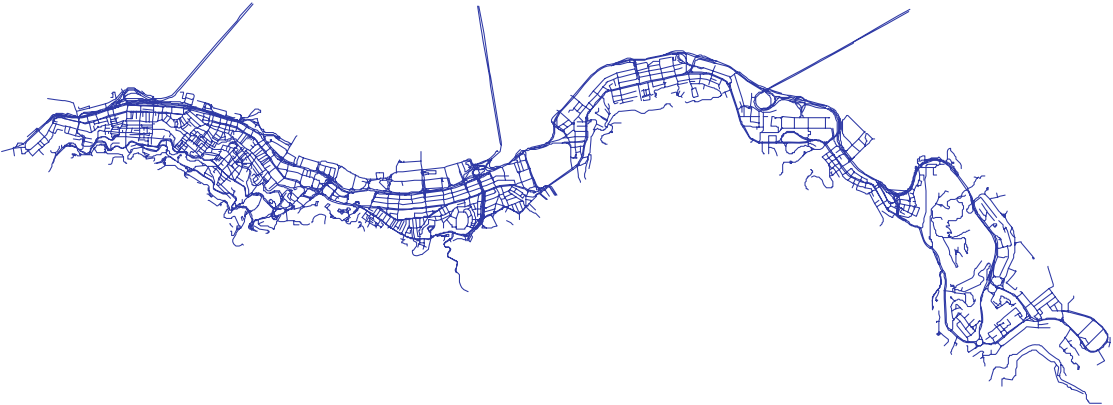
(j)

Appendix C: (color online) 9 sampled areas according to morphology

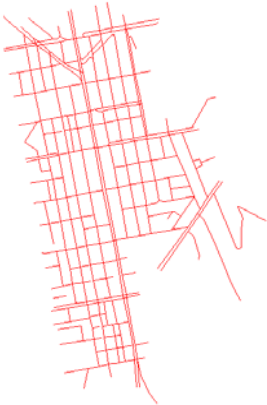
(a)

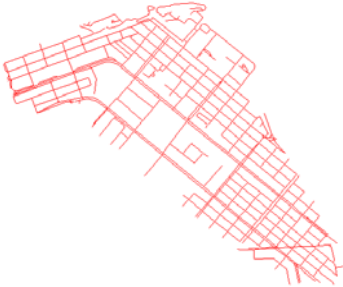
(b)

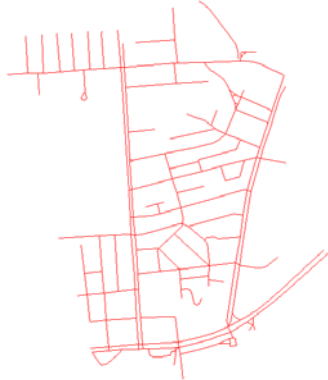
(c)

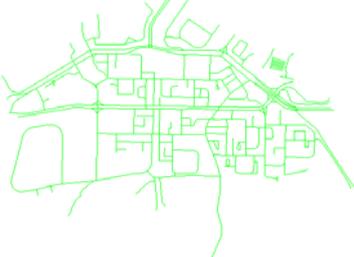
(d)

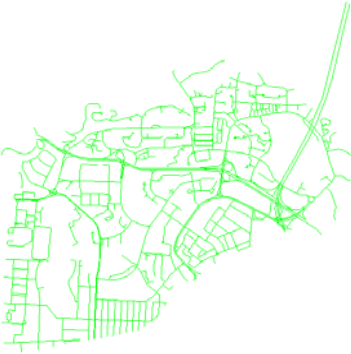
(e)

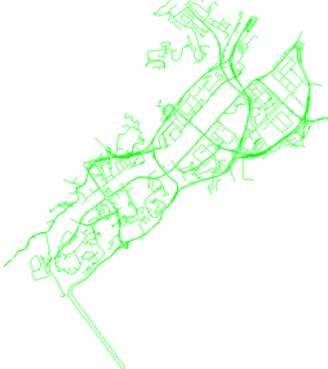
(f)

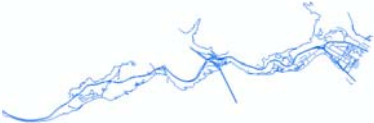
(g)

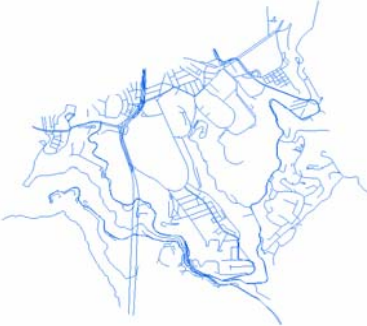
(h)

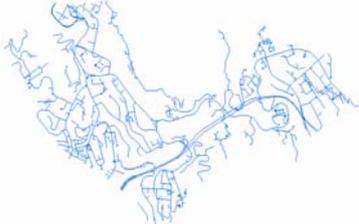
(i)